# Implied Correlation for Pricing multi-FX options

Pavel V. Shevchenko, CSIRO Mathematical and Information Sciences, Sydney, Australia

The number of multi-currency exotic options is large and growing. They naturally appeal to large international corporations who need to hedge their exposures in different currencies. Multi-currency options also attract sophisticated speculators having particular views on correlated movements of different Foreign Exchange Rates (FXRs). The most popular contracts include basket and barrier type options. It is common practice to estimate "true" option value via the expectation of the option payoff with respect to the so-called risk-neutral densities of the underlying assets.

Apparently, the value of a multi-currency option depends on correlations between underlying FXRs. In practice, it can be difficult to calculate correlations from historical returns because one has to decide on the length and frequency of time series and how to weight past returns. The estimates may not be stable, that is if the data are split into groups then the correlations for each group may differ significantly. Also, the option value should depend on correlations during the life of the option, which are predicted future correlations. Another possibility is to find implied correlations from market prices of contracts that estimate the market's perception of correlation. In this paper, we assume that there are liquid vanillas for all currencies involved in the multi-currency option and estimate correlations using implied volatilities of these vanillas. To price options written on several FXRs with the same denominating currency, financial practitioners and traders often use implied correlations calculated from implied volatilities of FXRs that form "currency triangles". However, some options may have underlying FXRs with different denominating currencies. In this paper, we present the formulas for the implied correlations between FXRs when denominating currencies are the same and different. The later one is not widely known. Using implied volatilities at different maturities, it is possible to approximate time dependent instantaneous correlations to price barrier options more accurately. Also, given arguments that objective and risk-neutral dependence structures are the same under certain general conditions (see e.g. J. V. Rosenberg, *Non-Parametric Pricing of Multivariate Contingent Claims*, The Journal of Derivatives 10 (3), pp.9-26, 2003) the implied correlations can be used in forecast analysis instead of correlations estimated from historical returns. The implied correlation can better estimate the predicted future correlation since it reacts quickly to market changes, see e.g. Campa, J.M. and Chang, P.H.K. (1998), *The Forecasting Ability of Correlations Implied in Foreign Exchange Options*, The Journal of International Money and Finance 17, pp.855-880.

**Model and Notation**

It is common in practice, for FX option valuation purposes, to assume that FXRs follow a multivariate geometric Brownian risk-neutral process

$$dX_{i/j} / X_{i/j} = (r_i(t) - r_j(t))dt + \sigma_{i/j}(t)dW_{i/j}$$
$$E[dW_{i/j} dW_{m/k}] = \rho_{i/j,m/k}(t)dt \quad (1)$$

Here, $X_{i/j}$ is the price of one unit of the *j*-th currency denominated in the *i*-th currency; $r_i(t)$ and $r_j(t)$ are instantaneous interest rates of the *i*-th and *j*-th currencies respectively; $\sigma_{i/j}(t)$ is instantaneous volatility; $W_{i/j}$ is the standard Wiener process; and $\rho_{i/j,m/k}(t)$ is the instantaneous correlation between different FXRs. Under the model (1), the price of a vanilla call(put) with strike *K* and maturity *T*, written on $X_{i/j}$, can be calculated using the well-known Garman and Kohlhagen (1983) formula (the modified Black and Scholes formula for the case of currency options)

$$call = \exp(-\hat{r}_i T) \times (F \times N[d_1] - K \times N[d_2]),$$
$$put = \exp(-\hat{r}_i T) \times (K \times N[-d_2] - F \times N[-d_1])$$
$$d_1 = (\ln\frac{F}{K} + \tfrac{1}{2}\hat{\sigma}^2_{i/j}T)/\hat{\sigma}_{i/j}\sqrt{T}, d_2 = d_1 - \hat{\sigma}_{i/j}\sqrt{T}$$

where $F = X_{i/j}\exp[(\hat{r}_i - \hat{r}_j)T]$ is the forward and $N[x]$ is standard Normal distribution function. The rates and volatility are simple integrals of the instantaneous rates and volatility: $\hat{r}_i = \int_0^T r_i(t)dt/T$, $\hat{r}_j = \int_0^T r_j(t)dt/T$ and $\hat{\sigma}_{i/j} = \hat{\sigma}_{i/j}(0,T) = (\int_0^T \sigma^2_{i/j}(t)dt/T)^{1/2}$ respectively. Given market vanilla prices the formula is inverted to calculate implied volatility $\hat{\sigma}_{i/j}$.

Let $Y_{i/j} = \ln[X_{i/j}(t=T)/X_{i/j}(t=0)]$ represent the log-increment over the time period *T*. The marginal distribution of $Y_{i/j}$ is Normal distribution with variance

$$Var(Y_{i/j}) = \hat{\sigma}^2_{i/j}T = \int_0^T \sigma^2_{i/j}(t)dt \qquad (2)$$

and the joint distribution of $Y_{i/j}$ and $Y_{m/k}$ is the bivariate Normal distribution with correlation coefficient

$$\hat{\rho}_{i/j,m/k} = \int_0^T \rho_{i/j,m/k}(t)\sigma_{i/j}(t)\sigma_{m/k}(t)dt / \hat{\sigma}_{i/j}\hat{\sigma}_{m/k}. \qquad (3)$$

**Implied correlation between FXRs with the same denominating currency**
Under model (1), the implied correlation between FXRs, $X_{i/k}$ and $X_{i/j}$, with the same denominating currency, i.e. $m=i$ in (1-3), can easily be found from the implied volatilities of these FXRs and the implied volatility of cross FXR, $X_{j/k}$, as follows. Consider three different currencies $i \neq j \neq k$ (e.g. $US, Euro, JPY). Assume that implied volatilities, $\hat{\sigma}_{i/j}$, $\hat{\sigma}_{i/k}$ and $\hat{\sigma}_{j/k}$ are available at maturity *T*. In the absence of arbitrage

$$X_{j/k} = X_{i/k}/X_{i/j} => Y_{j/k} = Y_{i/k} - Y_{i/j} \qquad (4)$$

Thus the variance of $Y_{j/k}$ can be written as

$$Var(Y_{j/k}) = Var(Y_{i/k}) + Var(Y_{i/j}) - 2\hat{\rho}_{i/k,i/j}\sqrt{Var(Y_{i/k})Var(Y_{i/j})} \qquad (5)$$

Substituting (2) in the above, i.e. assuming model (1), leads to the well-known formula for the implied correlation between $\ln X_{i/k}$ and $X_{j/k}$ via implied volatilities

$$\hat{\rho}_{i/k,i/j} = \frac{\hat{\sigma}_{i/k}^2 + \hat{\sigma}_{i/j}^2 - \hat{\sigma}_{j/k}^2}{2\hat{\sigma}_{i/k}\hat{\sigma}_{i/j}} \tag{6}$$

These correlations are commonly used to price multi-currency options where the payoff depends on several FXRs with the same denominating currency, e.g. basket options.

**Implied correlation between FXRs with different denominating currencies**

For the valuation of exotic options written on FXRs with different denominating currencies, the formula (6) cannot be used. This is the case, for example, for a barrier option on two FXRs, where one FXR determines how much the option is in or out of the money at maturity and the other FXR is related to the barrier. The underlying FXRs for this option may have different denominating currencies. This example is easily generalized to barrier options with more than two FXRs, i.e. some FXRs determine the payoff while other FXRs are knock-in/knock-out indicators. The FXRs linked to the barrier and FXRs in the payoff may have different denominating currencies. There are closed-form solutions for some barrier options in the case of two FXRs assuming constant parameters in (1). In the case of time-dependent parameters or more than two FXRs, numerical procedures are used to price barrier options. In either case, the correlation is required for option valuation. Below we derive the implied correlation between FXRs with different denominating currencies via implied volatilities of all related FXRs.

Consider four different currencies $i, j, k, m$ and find the implied correlation $\rho_{i/j,m/k}$ between $Y_{i/j}$ and $Y_{m/k}$ as follows. The variance of $Y_{i/j} + Y_{m/k}$ can be written as

$$Var(Y_{i/j} + Y_{m/k}) = (\hat{\sigma}_{i/j}^2 + \hat{\sigma}_{m/k}^2 + 2\hat{\rho}_{m/k,i/j}\hat{\sigma}_{i/j}\hat{\sigma}_{m/k})T \tag{7}$$

In the absence of arbitrage

$$X_{m/k} = X_{i/k} / X_{i/m} \Rightarrow Y_{m/k} = Y_{i/k} - Y_{i/m} \tag{8}$$

Thus, the formula (7) can also be written as

$$Var(Y_{i/j} + Y_{m/k}) = Var(Y_{i/j} + Y_{i/k} - Y_{i/m}) =$$
$$= T(\hat{\sigma}_{i/j}^2 + \hat{\sigma}_{i/k}^2 + \hat{\sigma}_{i/m}^2 + 2\hat{\rho}_{i/j,i/k}\hat{\sigma}_{i/j}\hat{\sigma}_{i/k} - 2\hat{\rho}_{i/j,i/m}\hat{\sigma}_{i/j}\hat{\sigma}_{i/m} - 2\hat{\rho}_{i/k,i/m}\hat{\sigma}_{i/k}\hat{\sigma}_{i/m}) \tag{9}$$

Calculating the correlations in the above equation using the "currency triangle" equation (6) and equating (7) and (9) we obtain

$$\hat{\rho}_{m/k,i/j} = \frac{\hat{\sigma}_{i/k}^2 + \hat{\sigma}_{m/j}^2 - \hat{\sigma}_{j/k}^2 - \hat{\sigma}_{i/m}^2}{2\hat{\sigma}_{i/j}\hat{\sigma}_{m/k}} \tag{10}$$

This equation allows us to find implied correlation between $X_{i/j}$ and $X_{m/k}$ using implied volatilities $\hat{\sigma}_{i/j}$ and $\hat{\sigma}_{m/k}$ of these FXRs and implied volatilities of cross FXRs $\hat{\sigma}_{i/k}$, $\hat{\sigma}_{m/j}$, $\hat{\sigma}_{j/k}$ and $\hat{\sigma}_{i/m}$. Equation (10) reduces to (6) if denominating currencies in $X_{i/j}$ and $X_{m/k}$ are the same, i.e. $m=i$, $\hat{\sigma}_{i/i}=0$.

**Discussion**

Equations (6, 10) are still valid for the relationship between implied correlation $\hat{\rho}(T_1,T_2)$ and implied volatilities $\hat{\sigma}(T_1,T_2)$ over future time period $[T_1,T_2]$ and is also valid for the relationship between instantaneous correlation $\rho(t)$ and instantaneous volatilities $\sigma(t)$. The value of a barrier option depends on the time-structure of volatilities and correlations. Using observed implied volatilities for different maturities $T_n$, $n=1,2,...$ the implied volatilities over future time periods $[T_n,T_{n+1}]$ can be found as usual via $\hat{\sigma}^2(T_n,T_{n+1})=[\hat{\sigma}^2(0,T_{n+1})T_{n+1}-\hat{\sigma}^2(0,T_n)T_n]/(T_{n+1}-T_n)$ and equations (6, 10) can be used to find implied correlations $\hat{\rho}(T_n,T_{n+1})$ over that periods. The instantaneous volatilities $\sigma_{i/j}(t)$ can be extracted from (2), assuming some functional form, and used in (10) to find instantaneous correlations $\rho_{i/j,m/k}(t)$. These can be used in Monte Carlo or other numerical procedures to get better estimates of barrier option prices. The easiest approach is to assume piecewise constant functions of time. Then instantaneous volatilities and correlations are just forward implied volatilities and correlations over corresponding time periods: $\rho_{i/j,m/k}(t)=\hat{\rho}_{i/j,m/k}(T_n,T_{n+1})$; $\sigma_{i/j}(t)=\hat{\sigma}_{i/j}(T_n,T_{n+1})$ for $t \in (T_n,T_{n+1}]$. Other functional forms for time-dependence can be used as well.

It is important to note that calculation correlations via implied volatilities allows to transfer correlation risk of the multi FX option into volatility risk in all underlying FXs. This is because option price becomes dependent on FX volatilities only when correlation is calculated by (6) or (10).

One of the consequences of model (1) is that the implied volatility is independent of strike, while the observed implied volatility usually exhibits strike dependence (the so-called volatility smile). Thus, in the presence of smile, the pricing of multi-asset options assuming model (1) and correlations (6, 10) cannot be consistent with all implied volatilities observed for different strikes. Under model (1), one can make pricing consistent with, for example, at-the-money (ATM) implied volatilities (for all available maturities) by using ATM volatilities to find instantaneous time dependent volatilities and correlations in the way described above. The pricing can be made consistent with volatility smiles of option underlying FXRs by considering more complicated models. For example, the multivariate extension of state-dependent volatility model where time dependent $\sigma_{i/j}(t)$ and $\rho_{i/j,m/k}(t)$ in model (1) are assumed to be time and state dependent functions $\sigma_{i/j}(X_{i/j},t)$ and $\rho_{i/j,m/k}(X,t)$ respectively. Although in this case $Var(Y_{i/j}) \neq \hat{\sigma}_{i/j}^2 T$, equation (10) is still valid for the relationship between instantaneous correlations and volatilities. Thus, after marginal calibration of $\sigma_{i/j}(X_{i/j},t)$ using observed implied volatilities, one can find instantaneous state-dependent correlations using (10). Solution and calibration of such model is quite complicated and is beyond the scope of this article.